# 422 Million Q Planar Integrated All-Waveguide Resonator with a 3.4 Billion Absorption Limited Q and Sub-MHz Linewidth

Matthew W. Puckett[1†], Kaikai Liu[2†], Nitesh Chauhan[2], Qiancheng Zhao[2], Naijun Jin[3], Haotian Cheng[3], Jianfeng Wu[1], Ryan O. Behunin[4], Peter T. Rakich[3], Karl D. Nelson[1], Daniel J. Blumenthal[2*]

[1]Honeywell International, Phoenix, AZ, USA
[2]Department of Electrical and Computer Engineering, University of California Santa Barbara, Santa Barbara, CA, USA
[3]Department of Applied Physics, Yale University, New Haven, CT, USA
[4]Department of Physics and Astronomy, Northern Arizona University, Flagstaff, AZ, USA
[†]Equal contribution authors
[*]Corresponding author (danb@ucsb.edu)
(Date 8/18/2020)

## ABSTRACT

High Q optical resonators are a key component for ultra-narrow linewidth lasers, frequency stabilization, precision spectroscopy and quantum applications. Integration of these resonators in a photonic waveguide wafer-scale platform is key to reducing their cost, size and power as well as sensitivity to environmental disturbances. However, to date, the intrinsic Q of integrated all-waveguide resonators has been relegated to below 150 Million. Here, we report an all-waveguide $Si_3N_4$ resonator with an intrinsic Q of 422 Million and a 3.4 Billion absorption loss limited Q. The resonator has a 453 kHz intrinsic linewidth and 906 kHz loaded linewidth, with a finesse of 3005. The corresponding linear loss of 0.060 dB/m is the lowest reported to date for an all-waveguide design with deposited upper cladding oxide. These are the highest intrinsic and absorption loss limited Q factors and lowest linewidth reported to date for a photonic integrated all-waveguide resonator. This level of performance is achieved through a careful reduction of scattering and absorption loss components. We quantify, simulate and measure the various loss contributions including scattering and absorption including surface-state dangling bonds that we believe are responsible in part for absorption. In addition to the ultra-high Q and narrow linewidth, the resonator has a large optical mode area and volume, both critical for ultra-low laser linewidths and ultra-stable, ultra-low frequency noise reference cavities. These results demonstrate the performance of bulk optic and etched resonators can be realized in a photonic integrated solution, paving the way towards photonic integration compatible Billion Q cavities for precision scientific systems and applications such as nonlinear optics, atomic clocks, quantum photonics and high-capacity fiber communications systems on-chip.

# INTRODUCTION

Ultra-high Q resonators play a critical role across a wide range of applications including ultra-narrow linewidth lasers[1–3], optical frequency combs[4–6], optical gyroscopes[7], optical atomic clocks[8] and quantum communications and computation[9–13]. These resonators, typically used for laser linewidth narrowing and frequency stabilization, have been relegated to benchtop and bulk-optic implementations. Record low 40 mHz laser linewidths and frequency stabilization of $1\times10^{-16}$ over 1 second have been achieved with a single crystal silicon cavity cryogenically cooled and environmentally isolated Fabry-Perot resonator[1] while table-top ultra-low expansion glass cavities can realize sub-Hz linewidth semiconductor lasers and frequency stabilization on the order of $2.7\times10^{-15}$ over 1 second[14]. Progress has been made with miniaturization of tapered-fiber and free-space coupled ultra-high Q bulk optical resonators[15–20] to achieve Qs of 63 Billion[18]. For example, state-of-the-art centimeter-scale microrod cavities with 1 Billion Q are capable of delivering a 25 Hz integral linewidth semiconductor laser with fractional frequency stability of $7\times10^{-13}$ at 20 ms in a compact centimeter structure[3].

Translating the performance of ultra-high Q resonators to integrated waveguide designs will lead to a dramatic reduction in size, power, cost and reduced sensitivity to environmental disturbances as well as enabling higher level of on-chip integration[21–23]. Designs that support high power linear operation through a large mode area and that mitigate thermo-optic frequency noise through a large resonator mode volume are desirable[24,25]. Integrated resonator Qs have been limited to below 100 Million for ring-based[26] and 150 Million for spiral-based[25] waveguide designs. Significant progress has been made with hybrid designs that employ an on-chip etched silica disc resonator, demonstrating intrinsic Qs of 206 Million[27] and recently 1.1 Billion Q[28]. However, these etched silica designs are not fully compatible with wafer-scale fabrication, are susceptible to environmental conditions, and need hermetic seal covering as well as careful mode engineering. The challenges to increasing the Q and reducing loss are dependent on reducing waveguide scattering losses with high-aspect ratio designs[29,30] and low surface roughness etching[31]. The ultimate limit is determined by material losses and eventually Rayleigh scattering[32–35]. A measure of the waveguide absorption-limited loss Q is a good metric for what is achievable for a given waveguide and resonator technology if waveguide scattering mechanisms can be mitigated[31]. Intrinsic loss sets the lower bound for the resonator full width half maximum (FWHM) linewidth. New solutions are needed for all-waveguide resonator designs with Qs approaching 500 Million, capable of exceeding several Billion, with sub-MHz FWHM resonances, large mode area and volume, and compatible with photonic integration and wafer-scale processing.

In this paper we report a significant advancement in integrated waveguide resonator performance. A $Si_3N_4$ bus-coupled ring-resonator with a measured intrinsic Q of 422 Million is demonstrated. The resonator has a 906 kHz full-width half maximum (FWHM) linewidth and a corresponding fineness of 3005, realizing a < MHz linewidth for the first time in a photonic integrated planar circuit. The intrinsic linewidth is 453 kHz and the corresponding linear loss of 0.060 dB m$^{-1}$ represents the lowest waveguide loss on-chip achieved with a deposited $SiO_2$ upper cladding[30]. Moreover, we report a 3.4 Billion absorption loss limited Q measured using a photothermal measurement technique[35,36]. These are the highest intrinsic and absorption loss limited Q factors and lowest linewidth reported to date for a photonic integrated resonator. This performance is achieved through a careful reduction of scattering and absorption loss components and redeposition of a thin nitride layer. By modeling scattering loss and measuring total intrinsic loss and absorption loss, we nail down each loss origin. In addition to these commonly known loss origins, we perform the secondary ion mass spectroscopy (SIMS) measurements to investigate the potential dangling bond resonances at the etched SiN/Oxide interface as another potential loss origin. The large resonator mode area and mode volume enables ultra-low laser linewidths and ultra-stable, ultra-low frequency noise reference cavities. These results demonstrate promise to bring the performance of bulk optic and etched resonators to planar all-waveguide solutions and pave the path towards integrated all-waveguide Billion Q cavities for atomic clocks, quantum computing and communications, precision spectroscopy and energy efficient coherent communications systems.

The low loss waveguide ring resonator is depicted in Fig. 1**a** and an SEM micrograph cross section is shown in Fig. 1**b**. The waveguide surface roughness couples the guided energy into radiation continuum causing scattering loss and the bulk material absorption converts the guided optical energy into heat leading to bulk absorption loss, as illustrated in Fig. 1**c**. Point defects on the waveguide surface created during the material deposition or waveguide etching processes can introduce coupling between different longitudinal modes or between the forward and backward propagating modes, causing random resonance splitting, as illustrated in Fig. 1**d**. Material deposition or waveguide etching processes can create reconstructed Si-Si bonds and dangling Si- and N- bonds, which can also become secondary bonds with hydrogen impurities such as Si-H, N-H, and Si-O-H. These defect bonds are a potential major origin of surface absorption loss[35,37]. Excess loss at bus-to-resonator coupling can be another major origin of resonator loss at ultra-high Q regime. It has been shown that careful design of bus-to-resonator coupling such as pulley coupling and weakly tapered coupling as opposed to straight coupling is less susceptible to introducing excess loss[26].

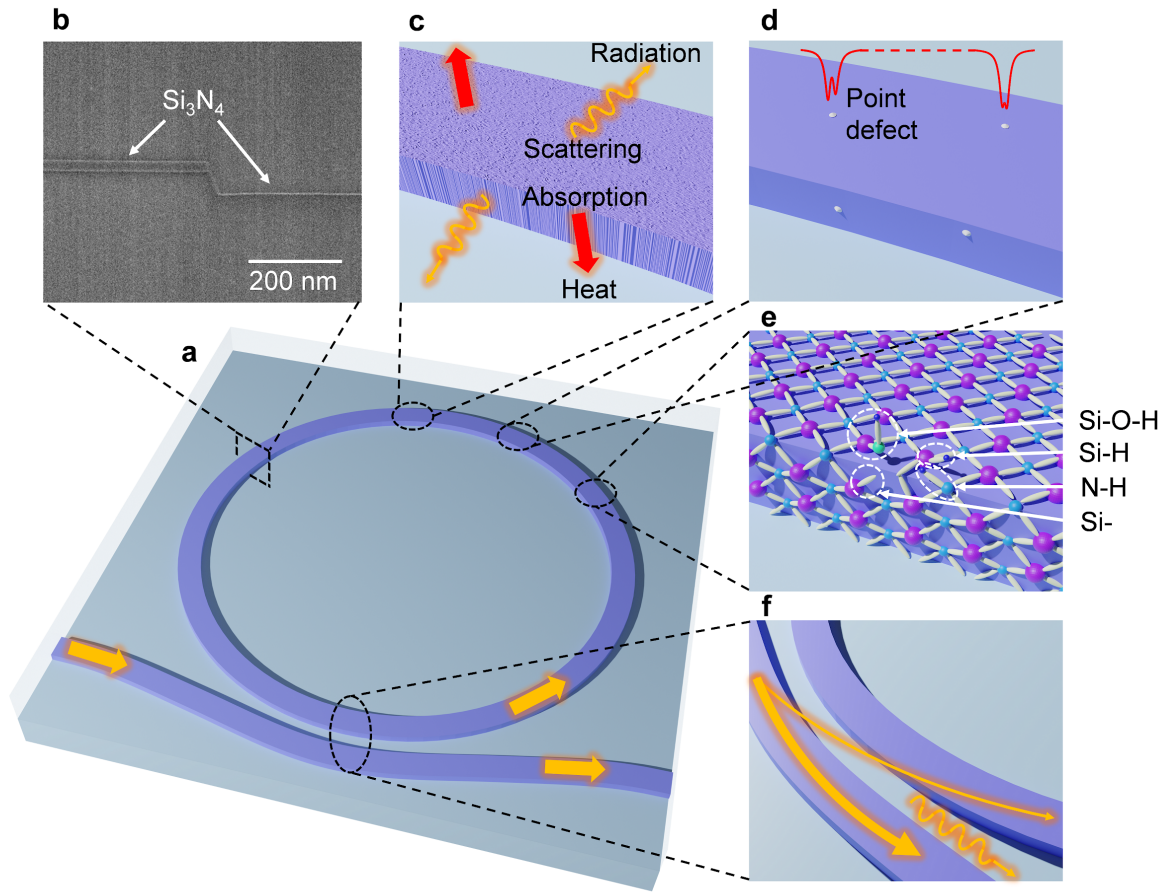

**Fig. 1. Ring resonator and loss mechanisms. a**, Overview of the resonator design. **b**, SEM image of the waveguide cross section. **c**, Waveguide surface roughness scatters the guide mode energy into radiation mode and bulk absorption generates heat. **d**, Point defects split resonances randomly. **e**, Defect bonds such as Si-O-H, Si-H, N-H, and dangling bonds cause surface absorption. **f**, Bus-to-resonator coupler scatters the guide energy into radiation mode and adds excess loss to the resonator.

# RESULTS

**Resonator design and fabrication:** The geometry of our $Si_3N_4$ waveguide resonator is designed to be 11 μm wide and 40 nm thick, and such a high-aspect-ratio geometry mitigates sidewall scattering[26,29]. The bus waveguide is designed to be 7 μm by 40 nm to ensure the single mode operation[29]. Our waveguide mode simulation suggests both the bus and resonator waveguides only support the fundamental TE mode and other higher order modes have large bending losses thus are not supported because of the asymmetric refractive indices in the upper and lower claddings, as demonstrated in the Supplementary Information. The resonator radius is 11.787 mm which is larger than the critical bending radius of the fundamental TE mode. The directional coupler is designed to be a weakly tapered coupler to avoid excess coupler loss[26]. Based on our previous measurement and coupling simulation, the gap between the bus and resonator waveguides is 6.898 μm such that the resonator is under-coupled, as discussed in the Supplementary Information. Since the high-aspect-ratio waveguide can suffer from the top and bottom surface roughness scattering, we include an additional nitride deposition step in our waveguide fabrication process through which the waveguide top surface can be potentially smoothed and the top surface scattering reduced.

The fabrication process flow includes standard wafer patterning, etching, upper cladding deposition, and annealing. The bottom cladding is 15 μm thick thermally grown oxide on the silicon substrate. The upper cladding is 6 μm TEOS-LPCVD deposited silicon dioxide. The final step is annealing at 1100 °C for 9 hours. The additional nitride deposition comes after etching and before upper cladding deposition: we deposit a few tens of nm of silicon nitride by LPCVD and subsequent heat treatment, creating a boundary between etched waveguide core and silicon dioxide which contributes to a smoother interface. The detailed fabrication process flow is illustrated in the Supplementary Information. The addition of the second SiN layer reveals the boundary between the upper and lower cladding in the scanning electron microscope (SEM) image of the cross section of the waveguide shown in Fig. 1**b**, which also allows us to observe that there is over-etching into the lower cladding by about 100 nm.

**Q-factor, linewidth and loss:** Spectral scans for the ultra-high-Q (UHQ) ring resonator with a tunable external cavity laser are performed for the Q factor characterization. Figure 3**a** shows the multi-FSR scan of our UHQ resonator fabricated with the boundary layer at 1550 nm. To demonstrate the loss reduction benefit of this technique, we fabricated another wafer's UHQ resonator devices (device under test, DUT) without the extra redeposition-and-annealing step for comparison ("control"). We employ a fiber Mach-Zehnder interferometer (MZI) with a calibrated FSR of 5.871 MHz for the optical frequency calibration. A Lorentzian fit of the resonance extracts the total linewidth $\gamma_T$, the loaded Q factor $Q_L = \omega_0/\gamma_T$, the coupling rate $\gamma_{ex}$, the intrinsic linewidth $\gamma_{in}$, and the intrinsic Q factor, $Q_{in} = \omega_0/\gamma_{in}$. The calibrated MZI linewidth measurement is performed on non-splitting resonances from 1550 nm to 1600 nm with intrinsic linewidth, intrinsic Q, and propagation loss for both the DUT and control shown in Fig. 2**c**, 2**d**, 2**e**, where we find the highest intrinsic Q of 422 Million at 1570 nm for the DUT. The spectral plots of the total linewidth, intrinsic linewidth, and coupling rate are shown in the Supplementary Information, which confirms the designed under-coupling operation at 1550 nm for the DUT. The linewidth measurement at 1570 nm shown in Fig. 2**a** reveals that the DUT reaches a total linewidth of 906 kHz and an intrinsic linewidth of 453 kHz. With the FSR at 1570 nm measured to be 2.720 GHz as shown in the supplementary, the corresponding finesse is 3005. To further confirm this highest Q, a ring-down experiment is performed at this resonance which gives an intrinsic Q of

434 Million. The comparison between the DUT and control shows that the additional nitride layer step significantly reduces the propagation loss.

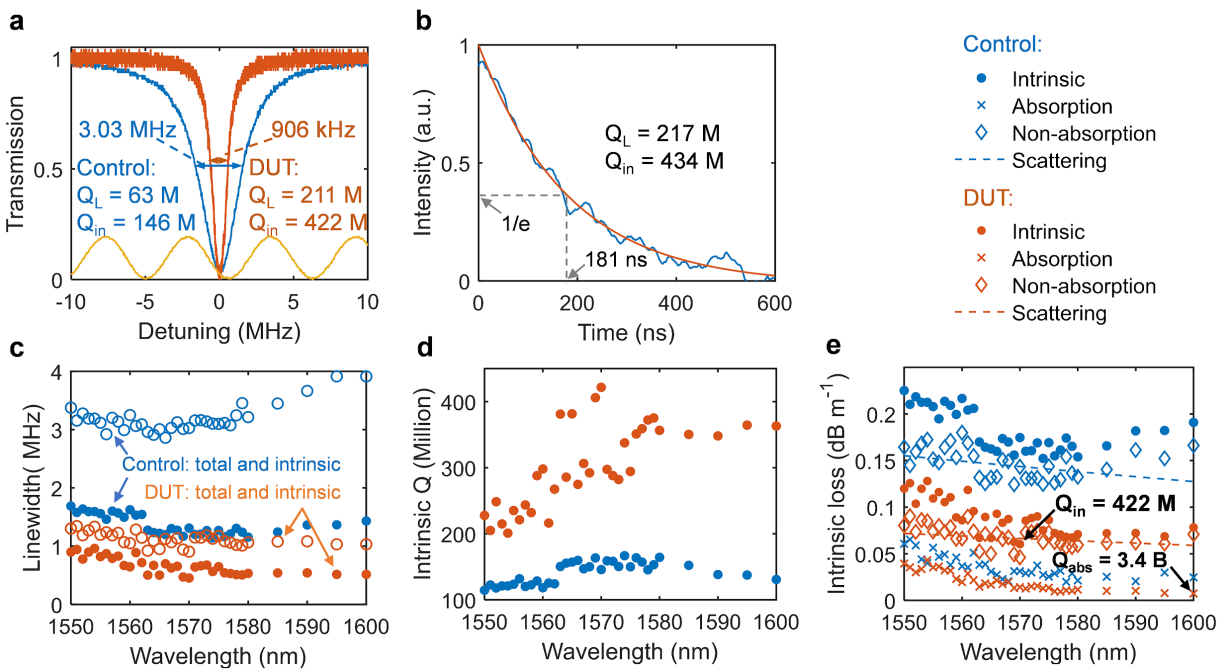

**Fig. 2. Linewidth and ring-down measurement of the DUT and control resonators, as well as the photo-thermal absorption loss measurement of the resonators. a**, Spectral scan for the fundamental TE mode of the control and DUT at 1570 nm. The total linewidth and intrinsic linewidth are extracted from a Lorentzian fit. The yellow sinusoidal signal is a fiber Mach-Zehnder interferometer with a calibrated FSR of 5.871 MHz (see the Supplementary Information about the calibration). M, million. **b**, Ring-down experiment for the resonance at 1570 nm of the DUT with the ringdown time $\tau$ and the loaded and intrinsic Q factors. **c**, Total and intrinsic linewidth spectrum, and **d**, intrinsic Q spectrum from 1550 nm to 1600 nm for both the DUT and control. **e**, Photo-thermal absorption loss measurement from 1550 nm to 1600 nm for both the DUT and control. The DUT resonator exhibits the highest intrinsic Q of 422 Million at 1570 nm and the absorption-limited Q of 3.4 Billion at 1600 nm. The non-absorption loss is fitted with the surface scattering loss model, indicated by dash lines..

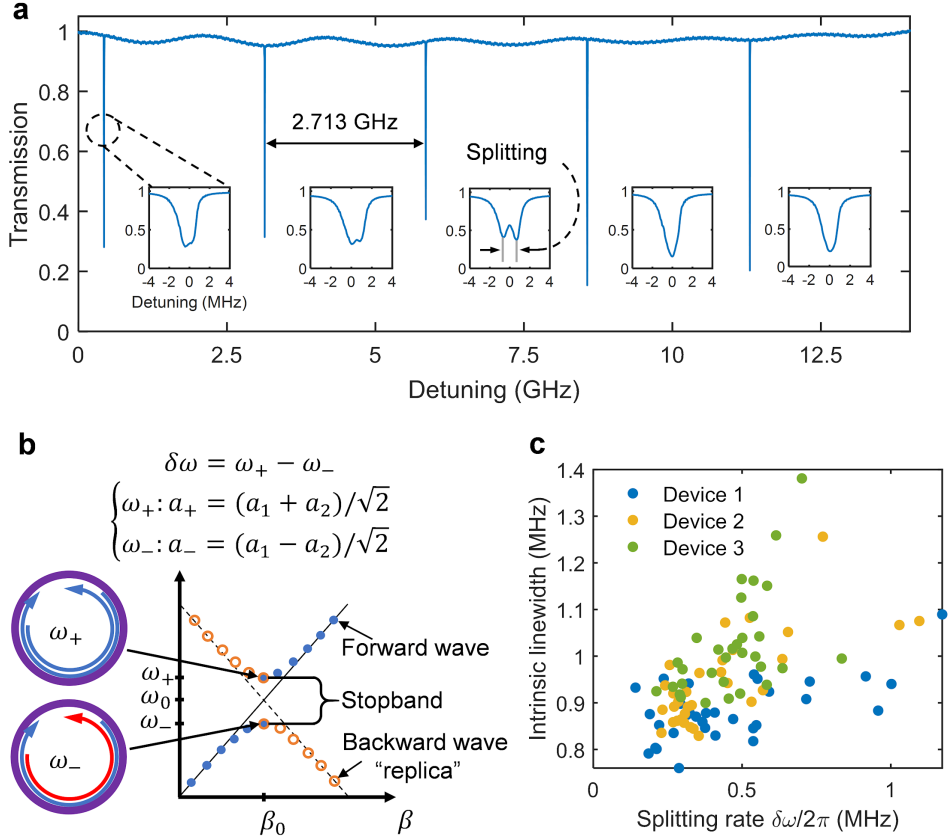

**Fig. 3. Resonance splitting. a**, Spectral scan across five FSRs at 1550 nm shows the single mode operation and the zoom-in views of each individual resonance reveals the random splitting at different resonances. **b**, The coupling between the forward and backward propagating waves creates a stopband resulting in resonance splitting. **c**, Scatter plot of the intrinsic linewidth versus the splitting rate near 1550 nm for three resonators fabricated on the same wafer.

**Resonance splitting:** Resonance splitting is often created intentionally and in a well-controlled manner by introducing coupling between the clockwise (CW) and counter-clockwise (CCW) modes of a ring resonator. This can be achieved, for example, by adding Bragg gratings to the waveguide[38,39], or by putting well-positioned scatterers near the resonator waveguide[40,41]. On the other hand, the mode coupling can be induced by waveguide roughness or surface defects resulting in resonance splitting, which can be purely random across different resonances. As illustrated in Fig. 3**b**, the coupling between the CW and CCW modes creates a stopband resulting in resonance splitting. Resonance splitting has been reported in high Q WGM resonators. Yet, resonance splitting has not been resolved and observed in planar waveguide resonators until the Q reaches 10 Million[31,36]. The coupled mode equation (CME) method models resonance splitting by incorporating the mode coupling[34,35,42,43], as demonstrated in the Supplementary Information. Resonance splitting also emerges in our UHQ waveguide resonator whereas it is not seen or not resolved in the control device. The insets in Fig. 3**a** reveal the zoom-in view of each individual resonance and different resonance splitting rates. The splitting appears random across different resonances. Moreover, the resonances that have less obvious splitting tend to have larger extinction ratios at resonance, which implies a narrower linewidth given that the resonator is under-coupled.

Figure 3c shows the plot of the intrinsic linewidth versus the splitting rate of 96 resonances near 1550 nm for three devices fabricated on one wafer, which shows a positive correlation between linewidth and splitting rate.

Resonance splitting has been utilized for suppression of higher order cascaded stimulated Brillouin scattering lasing[38] and nano-particle detection[40]. Yet, the random resonance splitting in the UHQ resonators could cast shadow on a lot of applications of UHQ waveguide resonators that prefer non-splitting resonances such as stimulated Brillouin lasers and Kerr frequency comb, especially when the linewidth is even narrower than the splitting.

**Absorption loss:** The Q-factor spectral characterization for the DUT shows a significant loss reduction of the smoothing step compared to the control. To fully understand its loss reduction benefit, we also perform photothermal absorption loss measurements. It has been demonstrated that the photothermal effect and consequent thermal bistability of a resonator can be characterized to measure absorption loss[35,36,44,45]. The photothermal effect emerges in the spectral scan across resonance when the on-chip power is large enough to induce a photothermal redshift of resonance that is comparable to or larger than the resonance linewidth and this provides information which can be used to extract the absorption loss as illustrated in the Supplementary Information. Hydrogen-impurity-related absorption has been identified as a significant loss origin in the ultra-low loss regime, which has an absorption peak near 1520 nm[30,36,44,45]. The recent broadband absorption loss measurement on ultra-low loss waveguides by Liu *et al.* showed that hydrogen-impurity-related absorption loss can in fact dominate the absorption loss near 1520 nm. In that work, the absorption-limited linewidth was reduced from 20 MHz for a partial anneal to around 0.3 MHz for a more complete anneal[44].

The measured absorption loss spectrum is shown along with intrinsic loss in Fig. 2e, where we find that the narrowest absorption-limited linewidth at 1600 nm from the device is 51 kHz corresponding to an absorption-limited Q of 3.4 Billion. It is worth noting that the scattering loss is still dominant in both the DUT and control. Now it is clear that the additional processing reduces both absorption loss and scattering loss. The absorption loss reduction could be from the high temperature anneal that helps drive out hydrogen and the scattering loss reduction could be from a smoother boundary between core and upper cladding created by the nitride layer. Based on the measured total intrinsic loss and absorption loss, we fit the non-absorption loss spectrum with our scattering loss model, as shown in the Supplementary Information. Since our scattering loss model suggests that the sidewall scattering loss for our high-aspect-ratio waveguide with a typical RMS sidewall roughness of 1 nm and a correlation length of 50 nm is on the order of 0.001 dB m$^{-1}$ and the top or bottom scattering loss with a typical RMS roughness of 0.2 nm and a correlation length of 10 nm is on the order of 0.1 dB m$^{-1}$, we take the sidewall scattering as negligible. The non-absorption loss is fitted with an effective top surface roughness of 0.24 nm and 0.35 nm for the UHQ and control devices, respectively (the correlation length is assumed to be a typical value of 10 nm[30,31,36]).

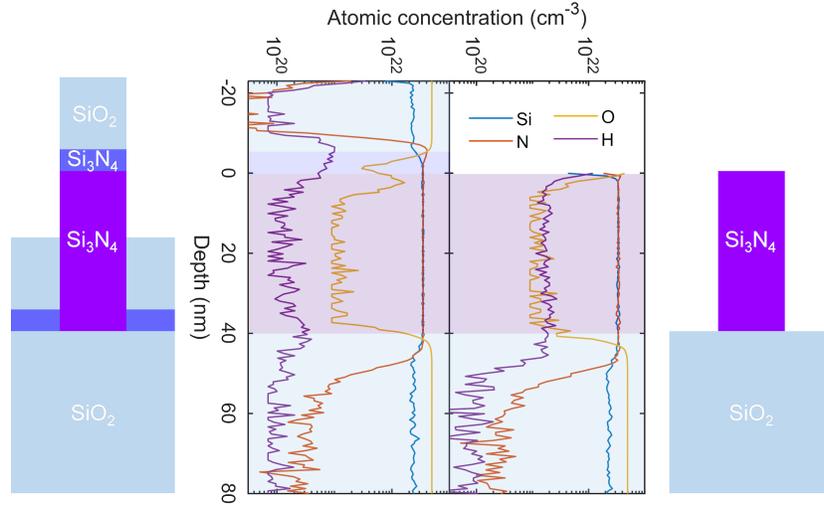

**Figure 4. Material characterization.** Secondary Ion Spectroscopy for hydrogen profiling for a wafer sample before the etching process and another wafer sample after the etching and nitride deposition.

**Surface loss effects:** The increasing trend of the absorption loss at shorter wavelengths shown in Fig. 3**e** suggests a likely hydrogen-impurity-related absorption peak near 1520 nm. Yet, it is not clear whether the hydrogen impurity is distributed in the bulk material or at the waveguide surface. To profile the densities of hydrogen and oxygen impurities, secondary ion mass spectroscopy (SIMS) is performed on a wafer sample after the etching process, and another wafer sample after the etching and nitride deposition and subsequent heat treatment, and the results are shown in Fig. 4. In the $Si_3N_4$ core, after the heat treatment, the hydrogen concentration drops to $1.3\times10^{20}$ (0.14 %) from $1.6\times10^{21}$ (1.7 %); at the boundary between the oxide and remaining redeposited $Si_3N_4$, there is a hydrogen concentration peak of $1.0\times10^{21}$ (1.1 %). Therefore, the heat treatment step also acts as an annealing process to drive out hydrogen which can contribute to the absorption loss reduction as we see in Fig. 2**e**. Surface absorption due to surface reconstruction could impose a Q-factor limit[35,37]. Here we notice that at the interfaces between different materials there is an increase of hydrogen which may be attached to reconstructed defect bonds such as Si-H, N-H, and Si-O-H, illustrated in Fig. 1**e**, which could impose a Q-factor limit on the order of Billion in the UHQ resonator performance.

## DISCUSSION

We report a significant advancement in integrated waveguide resonator performance with a measured intrinsic Q of 422 Million, an intrinsic linewidth of 453 kHz and corresponding linear loss of 0.060 dB m$^{-1}$, representing the lowest waveguide loss on-chip achieved with a deposited $SiO_2$ upper cladding and for the first time, < MHz linewidth for a waveguide photonic integrated resonator has been demonstrated. We have discussed how this performance is achieved through a careful reduction of scattering and absorption loss components and redeposition of a thin nitride layer and carried out detailed modeling of scattering loss as well as secondary ion mass spectroscopy (SIMS) measurements to investigate the potential dangling bond resonances at the etched SiN/Oxide interface as another potential loss origin. The large resonator mode area and volume are critical for ultra-low laser linewidths and ultra-stable, ultra-low frequency noise reference cavities. Additionally, we measure a 3.4 Billion absorption loss limited Q using a photothermal measurement technique, indicating that by carefully reducing the scattering losses, on-chip waveguide resonators exceeding 1 Billion Q are possible for future integrated applications. Other possible ways to increase the Q are through dual-polarization waveguides since the TM is less susceptible to waveguide roughness and has lower propagation loss. We anticipate that by a dual-polarization design of our waveguide in the future work could achieve even high loaded and intrinsic Q with TM mode. As we have noted that the annealing process drives out hydrogen, it could be beneficial to add an additional annealing process right after etching to drive out hydrogen in the first place. These results demonstrate promise to bring the performance of bulk optic and etched resonators to planar all-waveguide solutions and pave the path towards integrated all-waveguide Billion Q cavities for atomic clocks, quantum computing and communications, precision spectroscopy and energy efficient coherent communications systems.

## ACKNOWLEDGEMENT

This material is based upon work supported by the Defense Advanced Research Projects Agency (DARPA) and Space and Naval Warfare Systems Center Pacific (SSC Pacific) under Contract No. N66001-16-C-4017. The views and conclusions contained in this document are those of the authors and should not be interpreted as representing official policies of DARPA or the U.S. Government. We would like to thank Ron Polcawich and James Adleman for useful discussions. We also thank Jim Nohava, Joe Sexton, Jim Hunter, Dane Larson, Michael DeRubeis, and Jill Lindgren at Honeywell for their contributions to mask design and sample fabrication.

## DISCLAIMER

The authors declare no competing financial interest.

## CONTRIBUTIONS

M. W. P., K. L., and D. J. B. prepared the manuscript. M. W. P. conceived the resonator design and fabrication process. M. W. P., K. L., and N. C. tested the resonator devices with the calibrated MZI and ringdown. Q. Z. and J. W. fabricated the wafer samples for the SIMS testing. N. J. and H. C. developed and implemented the scattering loss modeling. K. L. and N. C. performed photothermal absorption loss measurements and analysis. All authors contributed to analyzing simulation and experimental results. D. J. B., K. D. N., and P. T. R. supervised and led the scientific collaboration.

## DATA AVAILABILITY

The data that support the plots within this paper and other findings of this study are available from the corresponding author on reasonable request.

# 422 Million Q Planar Integrated All-Waveguide Resonator with a 3.4 Billion Absorption Limited Q and Sub-MHz Linewidth: Supplementary Materials

Matthew W. Puckett[1†], Kaikai Liu[2†], Nitesh Chauhan[2], Qiancheng Zhao[2], Naijun Jin[3], Haotian Cheng[3], Jianfeng Wu[1], Ryan O. Behunin[4], Peter T. Rakich[3], Karl D. Nelson[1], Daniel J. Blumenthal[2*]

[1]Honeywell International, Phoenix, AZ, USA
[2]Department of Electrical and Computer Engineering, University of California Santa Barbara, Santa Barbara, CA, USA
[3]Department of Applied Physics, Yale University, New Haven, CT, USA
[4]Department of Physics and Astronomy, Northern Arizona University, Flagstaff, AZ, USA
[†]Equal contribution authors
[*]Corresponding author (danb@ucsb.edu)

(Date 8/18/2020)

**Fabrication flow and resonator design:** Figure S1 shows the fabrication process flow with the redeposition-and-anneal steps indicated as step 3 and 4.

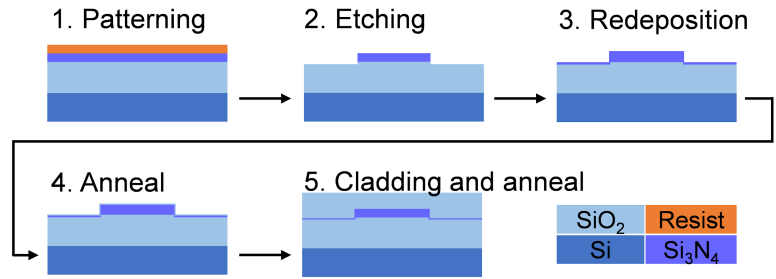

**Fig. S1. Fabrication process flow.**

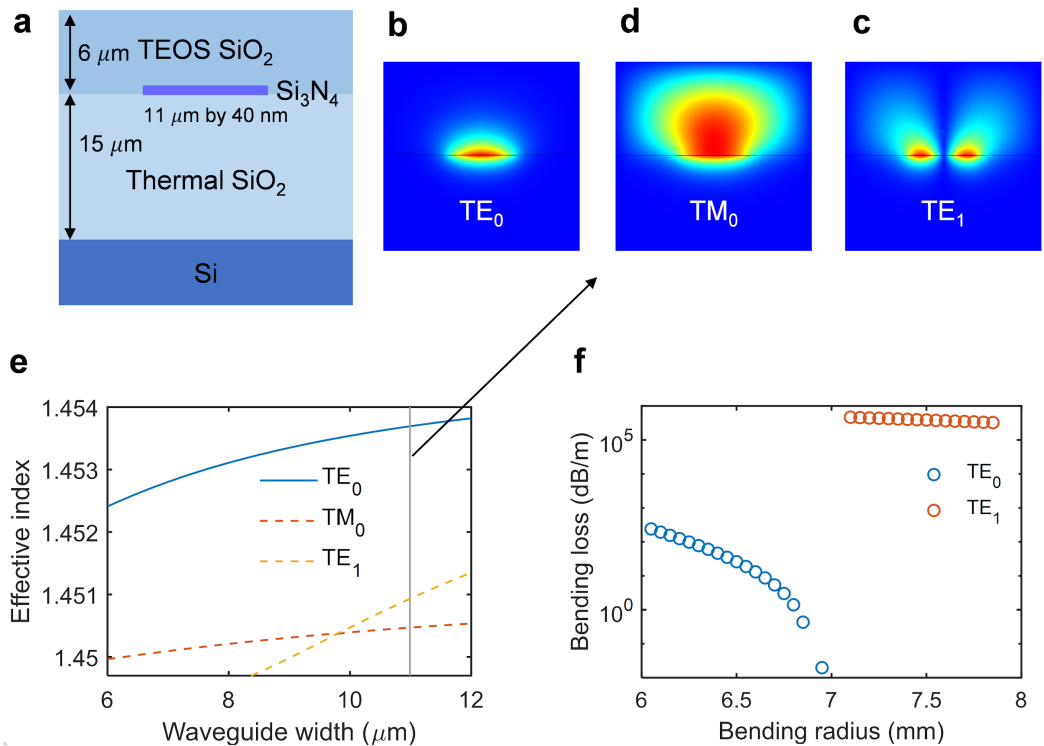

**Fig. S2. Single mode operation. a**, Cross-section diagram of the resonator waveguide. **b**, **c**, **d**, Mode profiles of the fundamental TE mode, leaky $TE_1$ mode, and leaky $TM_0$ mode. **e**, Effective index versus waveguide width with the waveguide thickness fixed at 40 nm. **c**, Single-mode operation in the ring resonator.

Using the measured refractive indices of the core and cladding materials that can be found in the Supplementary of our previous work[1], we perform the mode simulation with Lumerical to calculate the effective indices and the bending loss, as summarized in Fig. S2, which indicates both our bus and resonator waveguides only support the fundamental TE mode. In the bending radius range shown in Fig. S2**f**, the fundamental TM is not found in the mode solver and the fundamental TE mode has a critical bending radius of ~6.8 mm.

Since the Lorentzian fit of resonances to extract the intrinsic and coupling loss rates and the fitting does not distinguish one from another, we perform a Comsol simulation and a numerical calculation to fit the extracted coupling loss rate and to distinguish the intrinsic loss from the coupling loss[8]. Figure S3 summarizes the simulation and fitting for coupling loss from 1550 nm to 1600 nm, where Fig. S3**c** shows a good agreement between the coupling simulation and a direct measurement on a coupling testing structure. The testing device is on a stage the temperature of which is stabilized at the 1 mK temperature variance stability. Besides, the experiment of changing the temperature shows no clear temperature dependence of the coupling. The coupling is designed as a weakly tapered coupler to avoid any excess loss at the coupler, shown in a microscope image of the directional coupler in Fig. S3**b**. Should there be excess loss $\gamma$ besides the coupling coefficient $\kappa_c$ at the coupler, the resonator total linewidth is expressed as,

$$\gamma_{total} = \frac{c}{n_g L}(\alpha L + \kappa_c + \gamma)$$ (S1)

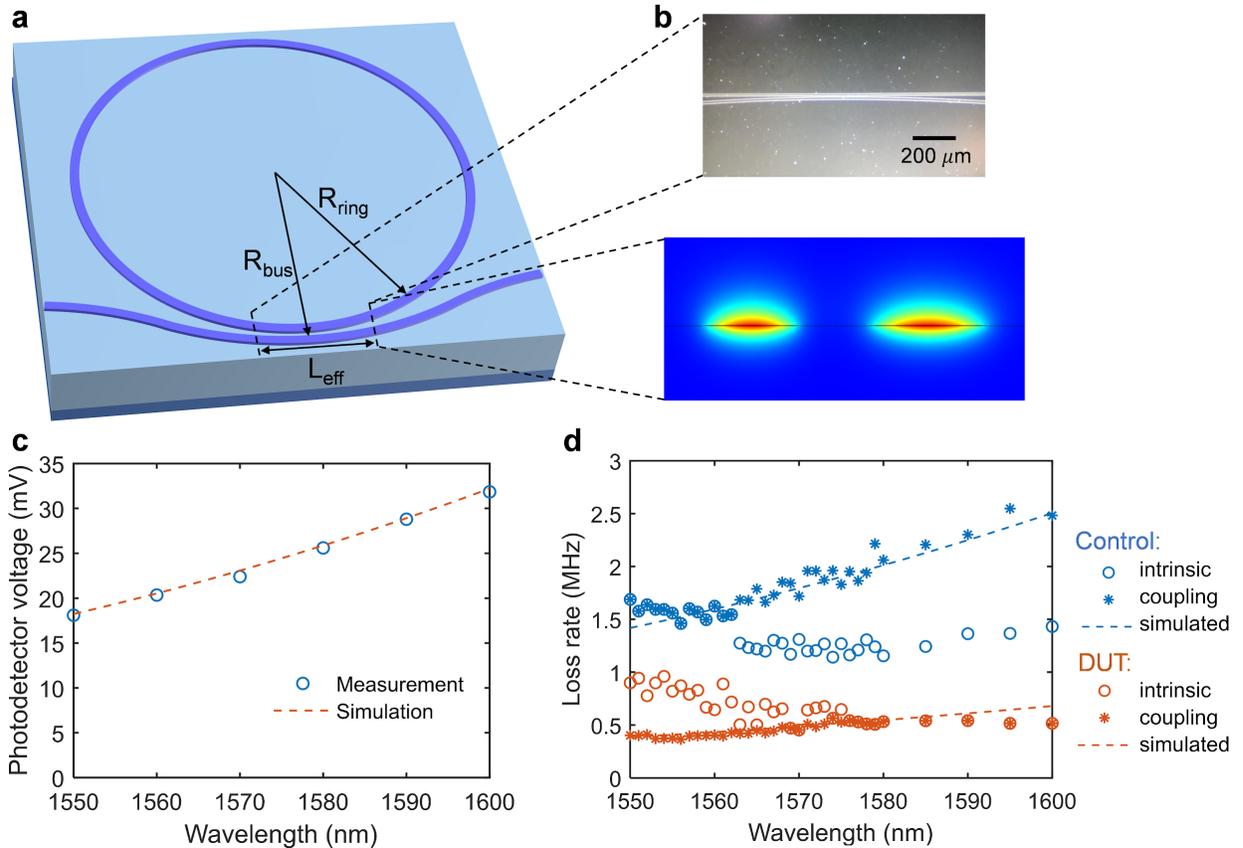

**Fig. S3. Weakly tapered coupling design, coupling simulation and measurement. a**, Weakly tapered coupling design. **b**, Microscope image of the coupler and mode simulation for the coupler cross section waveguides. **c**, Direct measurement of the coupling output on a photodetector along with the simulation fitting. **d**, Coupling simulation fitting of the coupling rate for both the DUT and control devices as shown by the dash lines.

**MZI calibration and ringdown measurement:** To calibrate the fiber Mach-Zehnder interferometer (MZI) free spectrum range (FSR), we employ an electro-optic modulator to add two sidebands the distance between which is defined by the modulation RF frequency, as illustrated in Fig. S4**a** and S4**b**. The FSR is calibrated to be 5.871 MHz. To further confirm the linewidth measurement with the MZI calibration, ringdown experiment is carried out at 1550 nm for both the DUT and control devices, and the ringdown results agree well with the linewidth measurements, as shown in Fig. S4.

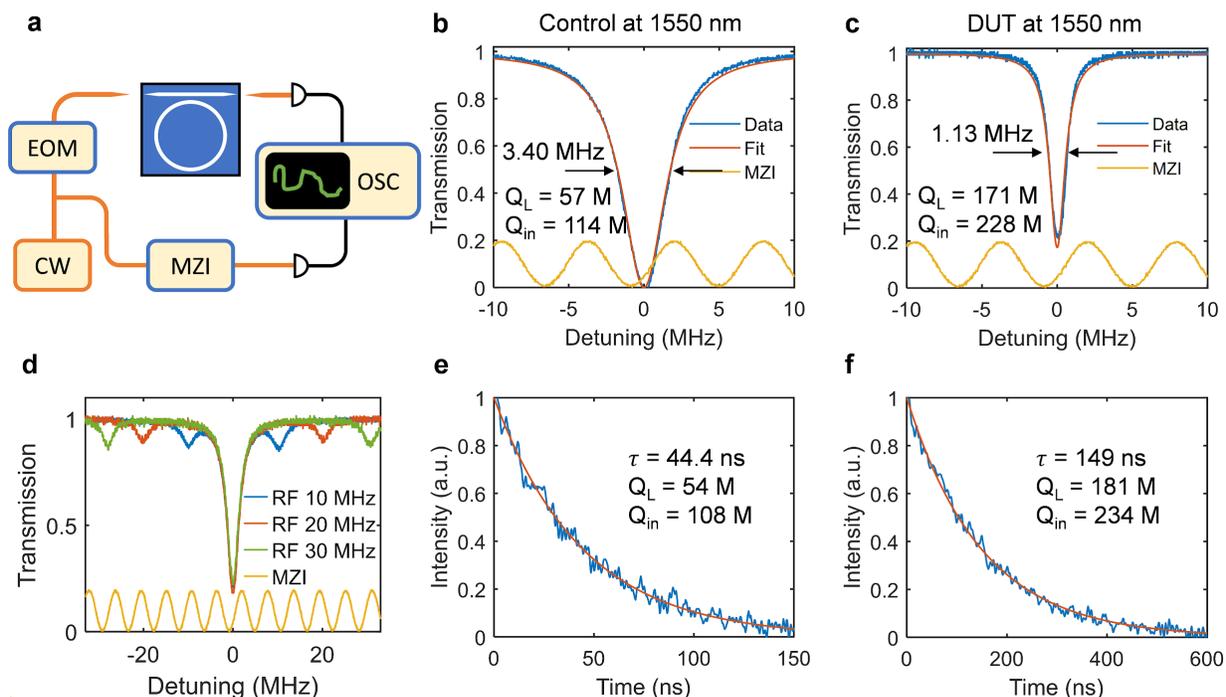

**Fig. S4. MZI calibration, calibrated MZI linewidth measurement, and ringdown experiment at 1550 nm. a**, Experiment setup diagram for MZI calibration and linewidth measurement. **d**, Electrooptic modulation sidebands for MZI FSR calibration. **b**, **c**, Calibrated MZI linewidth measurement at 1550 nm. **e**, **f**, Ringdown experiment at 1550 nm.

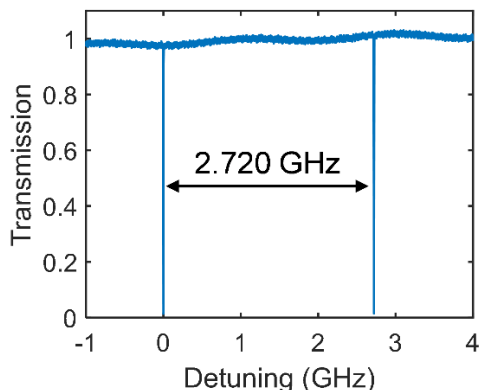

**Fig. S5. FSR at 1570 nm is measured to be 2.720 GHz.**

**Coupled mode equation for resonance splitting:** To describe the mode coupling between the clockwise (CW) and counter-clockwise (CCW) modes and the consequent resonance splitting, coupled mode equation (CME) method is widely used,

$$\frac{da_1}{dt} = -(i\Delta\omega + \frac{\gamma_T}{2})a_1 + i\frac{\kappa}{2}a_2 + i\sqrt{\gamma_{ex}}s_{in},$$
$$\frac{da_2}{dt} = -(i\Delta\omega + \frac{\gamma_T}{2})a_2 + i\frac{\kappa}{2}a_1,$$
$$s_{out} = s_{in} + i\sqrt{\gamma_{ex}}a_1,$$
(S2)

where $a_1$ and $a_2$ denote the CW and CCW modes, $s_{in}$ is the input mode, $\gamma_T = \gamma_{in} + \gamma_{ex}$ is the total loss including the intrinsic loss $\gamma_{in}$ and the external coupling loss $\gamma_{ex}$, the mode coupling coefficient is a complex number $\kappa = \kappa_R + i\kappa_I$, and $\Delta\omega$ is scan detuning. Solving the CME yields the doublet transmission lineshape where the mode coupling $\kappa = \kappa_R + i\kappa_I$, the splitting rate $\delta\omega$ and the linewidth difference $\delta\gamma$,

$$\left|\frac{s_{out}}{s_{in}}\right|^2 = \left|1 - \frac{\gamma_{ex}/2}{i(\Delta\omega - \kappa_R) + (\gamma_T - \kappa_I)/2} - \frac{\gamma_{ex}/2}{i(\Delta\omega + \kappa_R) + (\gamma_T + \kappa_I)/2}\right|^2, \quad (S3)$$

$$\delta\omega = 2\kappa_R, \delta\gamma = 2\kappa_I. \quad (S4)$$

Equation (S3) and (S4) are employed to fit the split resonances with $\kappa_I$ is set to be 0 to extract the intrinsic linewidth $\gamma_T = \gamma_{in} + \gamma_{ex}$, and the splitting rate $\delta\omega$.

**Scattering loss and mode coupling modeling:** The widely-used model to estimate waveguide scattering loss is the fully three dimensional volume current method (3D-VCM)[2–4], both scattering loss $\gamma_s$ and mode coupling $\kappa = \kappa_R + i\kappa_I$ can be estimated from the waveguide roughness profile. The far-field electric field produced by the roughness induced volume current and consequent coupling rate between the guide mode and radiation continuum are expressed as follows[4],

$$\boldsymbol{S}_m(\theta,\phi) = \sqrt{\frac{\omega^3}{16\pi^2\epsilon_0 c^3 U_m}} \int \Delta\epsilon(\boldsymbol{r}) \boldsymbol{E}_m(\boldsymbol{r}) \cdot (1 - \hat{\boldsymbol{k}}\hat{\boldsymbol{k}}) e^{-ik_0\hat{\boldsymbol{k}}\cdot\boldsymbol{r}} d^3\boldsymbol{r}, \quad (S5)$$

$$\Gamma_{m,m'} = \omega \int \boldsymbol{S}_m^*(\theta,\phi) \cdot \boldsymbol{S}_{m'}(\theta,\phi) \sin\theta d\theta d\phi, \quad (S6)$$

where $m$ denotes the CW and CCW modes, and $U_m$ is the mode energy in the waveguide, and $\Delta\epsilon(r)$ includes the roughness information. We can find that $\gamma_s = \Gamma_{11}$ and $\kappa_I = \Gamma_{12}$. With the first order perturbation theory, the coupling rate between the CW and CCW modes can be expressed as,

$$\kappa_R = \frac{\omega}{2U_m} \int \Delta\epsilon(\boldsymbol{r}) \boldsymbol{E}_m^*(\boldsymbol{r}) \cdot \boldsymbol{E}_{m'}(\boldsymbol{r}) d^3\boldsymbol{r}. \quad (S7)$$

The integrations in Eq. (S5-S7) incorporate the roughness information such as the sidewall roughness,

$$R_{side}(u_z) = \langle f_{side}(z) f_{side}(z+u_z)\rangle = \sigma_{side}^2 exp(-u_z/L_{side}), \quad (S8)$$

And top surface roughness,

$$R_{top}(u_x, u_y) = \langle f_{top}(x,y) f_{top}(x+u_x, y+u_y)\rangle = \sigma_{top}^2 exp[-(u_x+u_y)/L_{top}] \quad (S9)$$

Our model for estimating the scattering loss and mode coupling is validated by our model estimate getting the same sidewall scattering loss value from the same calculation carried out in these references[3,5]. Here we try a tentative estimation by choosing typical numbers for the waveguide roughness profiles: $\sigma_{side}$ = 2 nm, $L_{side}$ = 50 nm, $\sigma_{side}$ = 0.3 nm, and $L_{top}$ = 10 nm, which yields $\alpha_{top}$ = 0.101 dB m$^{-1}$, $\alpha_{side}$ = 0.001 dB m$^{-1}$, $\kappa_R = (2\pi)$ 0.632 MHz (from top roughness) + 0.001 MHz

(from sidewall roughness) = (2π) 0.633 MHz. These estimates are on the similar order of magnitude with the experimental measurements.

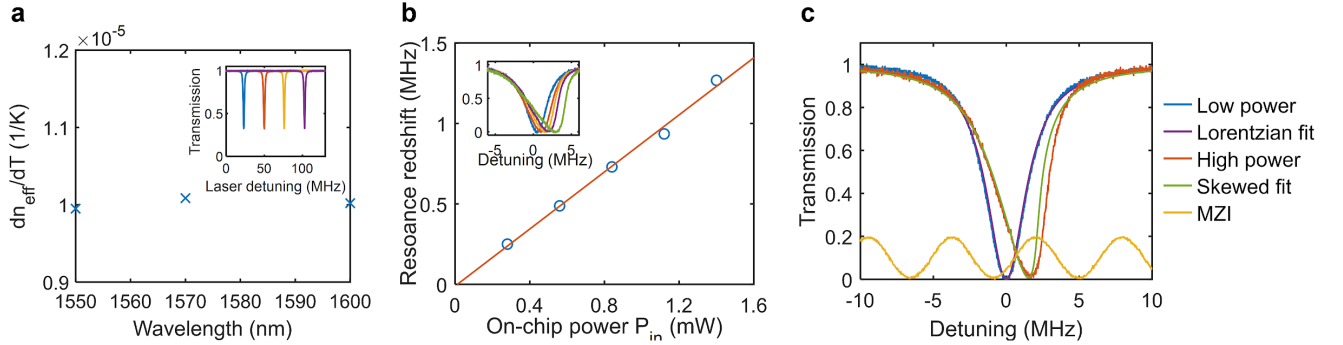

**Fig. S6. Photo-thermal heating, thermal bistability, absorption loss measurement. a**, Effective index change with respect to temperature shows no wavelength dependence. **b**, Photothermal effect is amplified by higher on-chip power. **c**, Normal Lorentzian fitting and skewed Lorentzian fitting extracts the intrinsic loss and absorption loss rates.

**Photothermal absorption loss measurement:** Transmission around resonance below 1 indicates the power dissipation in the resonator: $P_{disp} = P_{in}(1 - T_{res})$. Part of the dissipated power is absorbed and converted into heat: $P_{abs} = \xi P_{disp}$, where $\xi$ is absorption loss fraction and absorption loss rate can be expressed as $\gamma_{abs} = \xi \gamma_{in}$. Since only the waveguide is heated and the 1 mm thick Si substrate remains mostly undisturbed, thermal refractive effect dominate, and the thermal expansion effect is negligible. Using the thermo-optic coefficients of $SiO_2$ ($0.95 \times 10^{-5}$ 1/K) and SiN ($2.45 \times 10^{-5}$ 1/K) at 1550 nm reported in the literature[6,7], we perform a Comsol simulation that simulates the thermal heating due to absorption heating and estimates the redshift given an absorption power: $\delta f_{res} = \alpha P_{abs}$, illustrated in the inset of Fig. S6**b**. The simulation suggests $R_{th}$= 4.98 K/W, $\delta f_{res}/\delta T$= 1.23 GHz/K, and $\alpha = \delta f_{res}/P_{abs}$= 6.11 MHz/mW. In order to confirm the same thermo-optic coefficients at 1550 nm are valid for other wavelengths, we measure the resonance shift with a temperature increase and estimate the effective index change and Fig. S6**a** shows that there is not a clear wavelength dependence of the thermo-optic coefficients. A normal Lorentzian fit on the lower power transmission lineshape extracts the linewidths. With the extracted linewidths as the input parameters, we perform another fitting on the high power skewed lineshape with the following equation,

$$T = 1 - \frac{\gamma_{in}\gamma_{ex}}{[\Delta\omega - 2\pi f_D(1-T)]^2 + (\gamma_{in} + \gamma_{ex})^2/4}, \tag{S10}$$

where $f_D = \delta f_{res}/(1-T) = \xi \alpha P_{in}$ is the only parameter to be extracted and $\xi$ can be found. Figure S6**c** demonstrates the fitting processes.